# Chiral-field microwave antennas

## (Chiral microwave near fields for far-field radiation)

E. O. Kamenetskii, M. Berezin, and R. Shavit
Microwave Magnetic Laboratory
Department of Electrical and Computer Engineering, Ben-Gurion University of the Negev
Beer Sheva 84105, Israel
kmntsk@ee.bgu.ac.il; maksber@gmail.com; rshavit@ee.bgu.ac.il

*Abstract*—In a single-element structure we obtain a radiation pattern with a squint due to chiral microwave near fields originated from a magnetostatic-mode ferrite disk. At the magnetostatic resonances, one has strong subwavelength localization of energy of microwave radiation. Magnetostatic oscillations in a thin ferrite disk are characterized by unique topological properties: the Poynting-vector vortices and the field helicity. The chiral-topology near fields allow obtaining unique phase structure distribution for far-field microwave radiation.

*Keywords—antennas; ferrites; magnetostatic resonances; subwavelength energy localization, chiral fields*

## I. Introduction

In the search of novel communication systems, a strong interest arises in subwavelength structures with quasistatic oscillations as basic building blocks for controlling far-field electromagnetic radiation. Quasistatic resonances in small objects allow engineering of novel near-field structures with strong subwavelength localization of electromagnetic energy. These fields are distinguished by topological characteristics. Among them, there is a very unique property of the field chirality. In optics, the near field chirality was observed in metallic structures with plasmonic (electrostatic) resonances [1]. Recently, chiral near fields were obtained in microwaves based on small ferrite particles with magnetostatic resonances [2 – 6].

In this paper we show that microwave chiral fields originated from a small ferrite disk with magnetostatic oscillations can form far-field radiation pattern with a strong and controllable squint. Such an effect of a spatial mode division by a single radiation element is unique. It is very attractive for development of novel microwave antennas with controllable far-field phase structure distributions. It is well known that electromagnetic fields can carry not only energy but also angular momentum. The angular momentum is composed of spin angular momentum (SAM) and orbital angular momentum (OAM) describing its polarization state and the phase structure distribution, respectively. The research on OAM was not attractive until Allen *et al*. investigated the mechanism of OAM in 1992 [7]. Henceforth, more and more attention has been paid to OAM in both optical and radio domains. In contrast to SAM, which has only two possible states of left-handed and right-handed circular polarizations, the theoretical states of OAM are unlimited owing to its unique characteristics of spiral flow of electromagnetic energy [8]. Therefore, OAM has the potential to tremendously increase the spectral efficiency and capacity of communication systems [9]. Numerous experiments on OAM, originally in optical frequency and then in radio frequency, have been carried out.

The concept of OAM in radio frequency is relatively novel. For generating OAM in radio frequency a circular antenna array is used [10 – 12]. In this paper, we show, for the first time, that the far-field phase structure distribution for microwave radiation can be obtained by a single radiation element with chiral-topology near fields.

## II. Magnetostatic oscillations in small ferrite disks

Physical justification of multiresonance magnetostatic (MS) oscillations in microwave structures [] is based on the fact that in a small sample of a medium with strong temporal dispersion of the magnetic susceptibility, variation of electric energy is negligibly small and so the electric displacement current is negligibly small as well. In an analysis of such structures we should use three differential equations instead of the four-Maxwell-equation analysis of electromagnetic fields [13]:

$$\vec{\nabla} \cdot \vec{B} = 0 \quad (1)$$

$$\vec{\nabla} \times \vec{H} = 0 \quad (2)$$

$$\vec{\nabla} \times \vec{E} = -\frac{1}{c}\frac{\partial \vec{B}}{\partial t} \quad (3)$$

Taking into account a constitutive relation:

$$\vec{B} = \vec{H} + 4\pi\vec{m} \quad (4)$$

where *m* is the magnetization, one obtains from (1):

$$\vec{\nabla} \cdot \vec{H} = -4\pi \vec{\nabla} \cdot \vec{m} \quad (5)$$

This presumes an introduction of MS-potential wave functions ψ (*r*, *t*) for description of a magnetic field:

$$\vec{H} = -\vec{\nabla}\psi \qquad (6)$$

The spectral problem is formulated for MS-potential wave functions ψ (r, t), where a magnetization field is expressed as

$$\vec{m} = -\overset{\leftrightarrow}{\chi} \cdot \vec{\nabla}\psi \qquad (7)$$

Here $\overset{\leftrightarrow}{\chi}$ is the susceptibility tensor. Formally, in a system of (1) – (3), a potential magnetic field and a curl electric field should be considered as completely uncoupled fields. It turns out, however, that the magnetic and electric fields in (1) – (3) can be united. It was found that in a case of a quasi-2D ferrite disk, the spectral-problem solution for MS oscillations shows the presence of the unified (electric and magnetic) field structure which is different from the Maxwell-electrodynamics unified-field structure. We term the fields originated from the MS oscillations as magnetoelectric (ME) fields to distinguish them from regular electromagnetic (EM) fields [5]. MS oscillations in a quasi-2D ferrite disk are mesoscopically quantized states. Long range dipole-dipole correlation in position of electron spins in a ferromagnetic sample can be treated in terms of collective excitations of the system as a whole. The spectral solutions for the MS-potential wave function $\psi(\vec{r},t)$, has evident quantum-like attributes [2].

The incident EM wave has strong coupling with MS resonances of the ferrite disk and enable us to confine microwave radiation energy in subwavelength scales. In a vacuum subwavelength region abutting to a MS ferrite disk one can observe the quantized-state power-flow vortices [3 – 6]. The ME-field solutions give evidence for spontaneous symmetry breakings at the resonant states of MS oscillations. Because of rotations of localized field configurations in a fixed observer inertial frame, the linking between the EM and ME fields cause violation of the Lorentz symmetry of spacetime. In such a sense, ME fields can be considered as Lorentz-violating extension of the Maxwell equations [5, 6].

To characterize the ME-field singularities, the helicity (chirality) parameter was introduced. A time average helicity parameter for the near fields of a ferrite disk with MDM oscillations is defined as [5, 6, 14]:

$$F = \frac{1}{16\pi}\operatorname{Im}\left\{\vec{E}\cdot\left(\vec{\nabla}\times\vec{E}\right)^*\right\} \qquad (1)$$

One can also introduce a normalized helicity parameter, which shows a time-averaged space angle between rotating vectors $\vec{E}$ and $\vec{\nabla}\times\vec{E}$:

$$\cos\alpha = \frac{\operatorname{Im}\left\{\vec{E}\cdot\left(\vec{\nabla}\times\vec{E}\right)^*\right\}}{|\vec{E}||\vec{\nabla}\times\vec{E}|} \qquad (2)$$

In the regions where this parameter is not equal to zero, a space angle between the vectors $\vec{E}$ and $\vec{\nabla}\times\vec{E}$ is not equal to π/2. This breaks the field structure of Maxwell electrodynamics.

## III. MICROWAVE RADIATION STRUCTURE WITH A MS-MODE FERRITE DISK

As a radiation structure, we use a $TE_{10}$-mode rectangular X-band waveguide with a small hole in a wide wall. A hole diameter is 8 mm. A MS-mode ferrite disk is placed inside a waveguide symmetrically to its walls so that the disk axis is perpendicular to a wide wall of a waveguide. The disk axis coincide with a hole axis. The structure is shown in Fig. 1. The yttrium iron garnet (YIG) disk has a diameter of 3 mm and the disk thickness is 0.05 mm; the disk is normally magnetized by a bias magnetic field $H_0 = 4900$ Oe; the saturation magnetization of the ferrite is $4\pi M_s = 1880$ G. The sizes of a ferrite sample are extremely small compared to the EM wavelength in a waveguide.

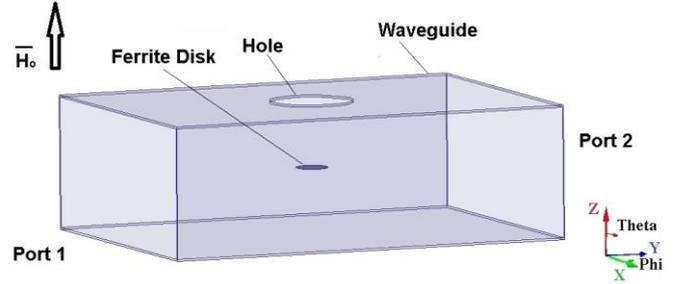

Fig.1. Waveguide radiation structure with a MS-mode ferrite disk and a hole in a wide wall.

In the shown microwave structure, a thin-film ferrite disk is the only resonant element. The multiresonance spectrum of MS oscillation in a closed waveguide structure with an embedded ferrite disk is well studied in our previous works [3 – 6, 14]. At the MS-mode-resonance frequencies one observes strong localization of microwave energy by a ferrite particle. For every resonance, the localized fields are topologically distinctive. One can observe power-flow vortices and non-zero helicity (chirality) parameters. Inside a waveguide, the power-flow vortices near a ferrite disk are non-symmetric with respect to waveguide walls. This non-symmetry, being different for different directions of the wave propagation in a waveguide, does not affect practically, however, on nonreciprocity in the microwave propagation from one port to another. Figs. 2 (a) and (b) show numerical results of the Poynting-vector distributions inside a waveguide near a ferrite disk at the MS-resonance frequency of 8.146GHz. One can see strong energy localization and non-symmetrically distinguishable pictures of the power-flow vortices for different directions of the wave propagation in a waveguide. At the same time, at the frequency beyond the MS resonance, the disk slightly perturbs the waveguide field [see Fig 2 (c)].

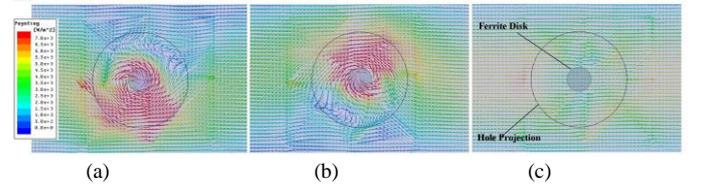

(a)        (b)        (c)

Fig.2. The Poynting-vector distributions inside a waveguide near a ferrite disk. (a) wave propagation from port 1 to port 2 at a MS resonance; (b) wave propagation from port 2 to port 1 at a MS resonance; (c) wave propagation at frequency beyond a MS resonance. The MS-resonance frequency is 8.146GHz.

When a waveguide has a hole (see Fig. 1), nonsymmetrical pictures of the power-flow distribution near a hole are evident at the MS-mode resonances. For different directions of the wave propagation in a waveguide, one has different pictures of the power-flow distribution near a hole. Fig. 3 shows numerical results of the Poynting-vector distribution in a small vacuum cylinder above a waveguide hall. Figs. 3 (a) and (b) correspond to two opposite directions of the wave propagation in a waveguide at a MS-mode resonance frequency of 8.146GHz. Evidently, there are non-symmetrical pictures with respect to the waveguide axes. When the frequency is beyond the MS-mode resonances, one has symmetry with respect to the waveguide axes. Fig. 3 (c) shows such a symmetrical picture of the power-flow distribution near a hole.

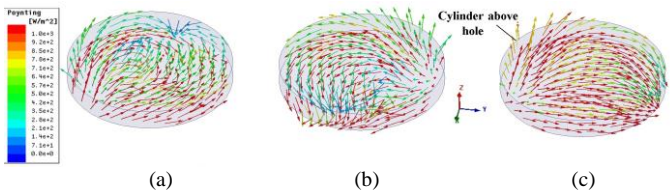

(a)  (b)  (c)

Fig.3. The Poynting-vector distributions above a hole. (a) wave propagation from port 1 to port 2 at a MS resonance; (b) wave propagation from port 2 to port 1 at a MS resonance; (c) wave propagation at frequency beyond a MS resonance. The MS-resonance frequency is 8.146GHz.

The symmetry-breaking properties of the fields originated from a ferrite disk at the resonant states of MS oscillations are exhibited not only by peculiar power-flow distributions. As very important characteristics, there are also distributions of the field helicity (chirality) parameter. Fig. 4 shows the normalized helicity parameter of the field above a hole in the XY plane (see Fig. 1). Figs. 4 (a) and (b) correspond to two opposite directions of the wave propagation in a waveguide at a MS-mode resonance frequency of 8.146GHz. Fig. 4 (c) shows that at a frequenciy beyond the MS resonance the fields are without any symmetry breakings and so no helicity properties are observed.

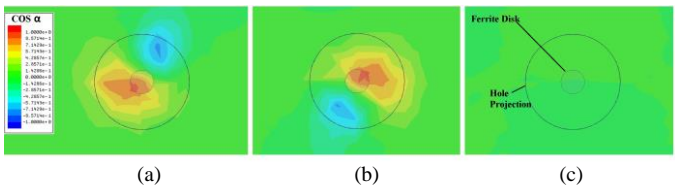

(a)  (b)  (c)

Fig.4. The normalized helicity parameter of the field above a hole. (a) wave propagation from port 1 to port 2 at a MS resonance; (b) wave propagation from port 2 to port 1 at a MS resonance; (c) wave propagation at frequency beyond a MS resonance. The MS-resonance frequency is 8.146GHz.

As an additional characterization of the symmetry breaking property of the ME fields, we show in Fig. 5 the normalized helicity parameter of the field in the YZ-plane cross section of the structure in Fig. 1 in the regions both inside and outside a waveguide.

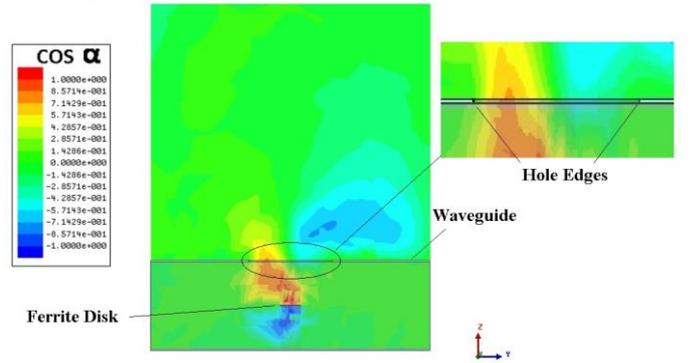

Fig.5. The normalized helicity parameter of the field in the YZ-plane cross section of the structure in the regions both inside and outside a waveguide at the MS-resonance frequency 8.146GHz. The wave propagates from port 1 to port 2.

## IV. RADIATION PATTERNS

While the shown non-symmetry in the Poynting-vector distributions does not affect on nonreciprocity between the waveguide ports, this gives strong nonreciprocity in the radiation patterns. It is well known that a standard single radiation element does not exhibit a radiation pattern with a squint. To obtain a squint in the radiation pattern, an array of radiation elements has to be used. In our single-element structure we obtain a radiation pattern with a squint due to chiral microwave near fields originated from a MS-mode ferrite disk. Fig. 6 shows the single-element radiation patterns for two cut planes, $\varphi = 0°$ and $\varphi = 90°$, at frequency 8.146GHz.

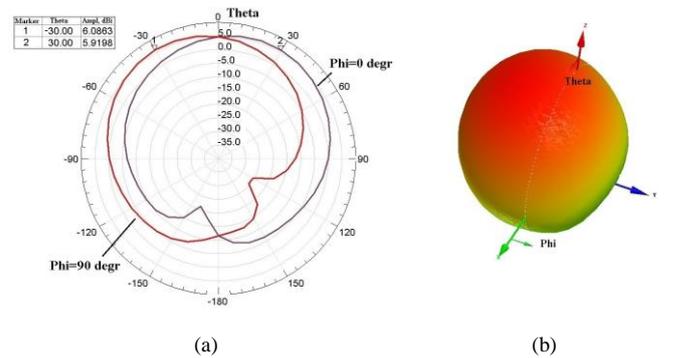

(a)  (b)

Fig.6. Single-element radiation patterns for two cut planes, $\varphi = 0°$ and $\varphi = 90°$, at the MS-resonance frequency 8.146GHz. The wave in a waveguide propagates from port 1 to port 2. (a) 2-D pattern; (b) 3-D pattern.

In Fig. 7, we show the radiation patterns ($\varphi = 0°$) for two different directions of the wave propagation in a waveguide at the MS-resonance frequency. We compare these patterns with a radiation pattern of a structure at a frequency beyond the MS-mode resonance.

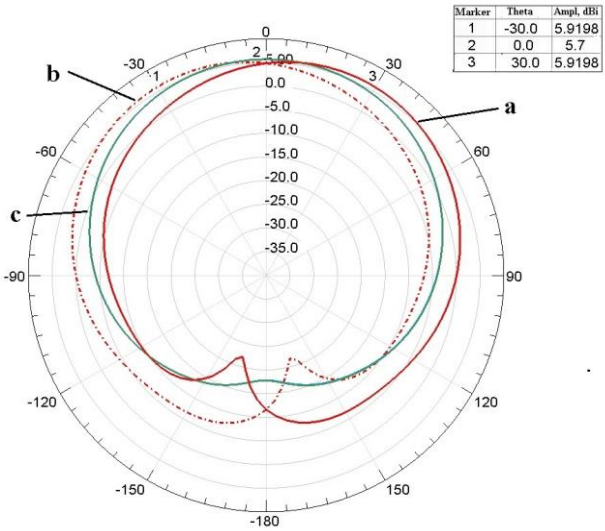

Fig.7. Radiation patterns ($\varphi = 0°$) for different directions of the wave propagation in a waveguide. (a) wave propagation from port 1 to port 2 at a MS resonance; (b) wave propagation from port 2 to port 1 at a MS resonance; (c) wave propagation at frequency beyond a MS resonance. The MS-resonance frequency is 8.146GHz.

Our preliminary results show also that the desired squint angles can be obtained by variation of a hole diameter and also by use of an elliptical-form hole.

## V. CONCLUSION

In this paper we showed that microwave chiral fields originated from a small ferrite disk with magnetostatic oscillations can form far-field radiation pattern with a strong and controllable squint. To the best of our knowledge, this is the first time demonstration of a radiation pattern with a squint obtained by a single radiation element. Such an effect of a spatial mode division by a single radiation element is unique could be attractive for development of novel microwave antennas with controllable phase structure distributions.